\documentclass[aip,groupedaddress,superscriptaddress,reprint]{revtex4-1}
\usepackage[bookmarks=true,colorlinks,linkcolor=OrangeRed,urlcolor=NavyBlue,citecolor=RoyalBlue]{hyperref}
\usepackage{amsmath,amssymb,color,comment,physics}
\usepackage[makeroom]{cancel}
\usepackage[caption=false]{subfig}
\usepackage{mathrsfs}
\usepackage{graphicx}
\usepackage{subfig}
\usepackage[countmax]{subfloat}
\usepackage[english]{babel}
\usepackage[dvipsnames]{xcolor}
\usepackage{mathtools}

\usepackage{ulem}
\DeclareUnicodeCharacter{2212}{\textendash}

\usepackage{braket}

\definecolor{mypurple}{rgb}{0.49,0.18,0.56}
\definecolor{mygold}{rgb}{0.93,0.49,0.13}
\definecolor{mygreen}{rgb}{0,0.5,0}
\definecolor{myblue}{rgb}{0,0,0.75}
\definecolor{mymagenta}{cmyk}{0,1,0,0.12}
\definecolor{mygray}{rgb}{0.5,0.5,0.5}

\usepackage{umoline}

\definecolor{mypink1}{rgb}{0.858, 0.188, 0.478}

\begin{document}

\title{Generalized Discrete Truncated Wigner Approximation for Nonadiabtic Quantum-Classical Dynamics}

\author{Haifeng Lang}
\affiliation{Theoretical Chemistry, Institute of Physical Chemistry, Heidelberg University, Im Neuenheimer Feld 229, 69120 Heidelberg, Germany }
\affiliation{INO-CNR BEC Center and Department of Physics, University of Trento, Via Sommarive 14, I-38123 Trento, Italy}

\author{Oriol Vendrell}
\affiliation{Theoretical Chemistry, Institute of Physical Chemistry, Heidelberg University, Im Neuenheimer Feld 229, 69120 Heidelberg, Germany }

\author{Philipp Hauke}
\affiliation{INO-CNR BEC Center and Department of Physics, University of Trento, Via Sommarive 14, I-38123 Trento, Italy} 





\begin{abstract}
Nonadiabatic molecular dynamics occur in a wide range of chemical
reactions and femtochemistry experiments involving electronically
excited states. These dynamics are hard to treat numerically as
the system's complexity increases and it is thus desirable to
have accurate yet affordable methods for their simulation.
Here, we introduce a linearized semiclassical method, the generalized discrete truncated Wigner approximation (GDTWA), which is well-established in the context of quantum spin lattice systems, into the arena of chemical nonadiabatic systems. In contrast to traditional continuous mapping approaches, e.g. the Meyer–Miller–Stock–Thoss and the spin mappings, GDTWA samples the electron degrees of freedom in a discrete phase space, and thus forbids an unphysical unbounded growth of electronic state populations. 
The discrete sampling also accounts for an effective reduced but non-vanishing zero-point energy without an explicit 
parameter, which makes it possible to treat the identity operator and other operators on an equal footing. 
As numerical benchmarks on two Linear Vibronic Coupling models show, GDTWA has a satisfactory accuracy in a wide parameter regime, independently of whether the dynamics is dominated by relaxation or by coherent interactions.
Our results suggest that the method can be very adequate to treat challenging nonadiabatic dynamics problems in chemistry and related fields. 

\end{abstract}

\pacs{Valid PACS appear here}
\maketitle
\section{Introduction}

The phase space representation is a powerful tool for computing quantum dynamics, with various linearized approximation methods 
having been developed by diverse  communities over the years, from quantum chemists to physicists.  \cite{hillery1984distribution,steel1998dynamical,blakie2008dynamics,polkovnikov2010phase,schachenmayer2015many,zhu2019generalized,davidson2015s,wurtz2018cluster,polkovnikov2003quantum,orioli2017nonequilibrium,pucci2016simulation,meyera1979classical,cotton2013symmetrical,stock1997semiclassical,cotton2013symmetrical2,liu2017isomorphism,he2019new,liu2016unified,miller2017classical,saller2019identity,saller2019improved,sun1998semiclassical,kim2008quantum,kelly2012mapping,huo2011communication,huo2013communication,huo2012consistent,hsieh2012nonadiabatic,hsieh2013analysis,kapral1999mixed,stock1999flow,muller1999flow,cotton2015symmetrical,meyer1979classical,mannouch2020partially,mannouch2020partially2,runeson2019spin,runeson2020generalized}.
Physicists often subsume those methods under the name of Truncated Wigner Approximations (TWA) with many family members \cite{hillery1984distribution,steel1998dynamical,blakie2008dynamics,polkovnikov2010phase,schachenmayer2015many,zhu2019generalized,davidson2015s,wurtz2018cluster,polkovnikov2003quantum,orioli2017nonequilibrium,pucci2016simulation}, whereas chemists usually call them mapping approaches, including the Meyer–Miller–Stock–Thoss (MMST) mapping \cite{meyera1979classical,cotton2013symmetrical,stock1997semiclassical,cotton2013symmetrical2,liu2017isomorphism,he2019new,liu2016unified,miller2017classical,saller2019identity,saller2019improved,sun1998semiclassical,kim2008quantum,kelly2012mapping,huo2011communication,huo2013communication,hsieh2012nonadiabatic,hsieh2013analysis,kapral1999mixed,stock1999flow,muller1999flow,huo2012consistent} and spin mapping (SM) \cite{cotton2015symmetrical,meyer1979classical,mannouch2020partially,mannouch2020partially2,runeson2019spin,runeson2020generalized}. The key idea of these methods is to sample the quantum distribution of the initial states as the Wigner quasiprobability distribution, and neglect higher-order quantum corrections of the Moyal bracket, thus rendering the evolution equations classical. One of the most important reason researchers are interested in these approaches is that the simulations using the classical dynamics are computationally cheap and the Monte Carlo sampling is trivially parallelizable. Hence, they can be applied to large systems, which is usually impossible for the numerically exact full quantum dynamics \cite{polkovnikov2010phase,zhu2019generalized}. 
Higher-order quantum corrections can also be introduced systematically\cite{polkovnikov2010phase,polkovnikov2003quantum,hsieh2013analysis,huo2012consistent}. These approaches are exact in the classical limit and the noninteracting limit. They can also provide reliable qualitatively correct results for short time dynamics when the system is not far away from the classical limit, and it is possible to capture the long-time detailed-balance behavior \cite{bellonzi2016assessment} or hydrodynamic phenomena \cite{zhu2019generalized,wurtz2018cluster,wurtz2020quantum} for specific models. Typical interesting systems that are suitable for these approaches include models from quantum optics \cite{gardiner2004quantum,walls2007quantum}, cold atoms \cite{ruostekoski2005dissipative,isella2005nonadiabatic,scott2009quantifying}, quantum spin chains \cite{schachenmayer2015many,zhu2019generalized,orioli2017nonequilibrium,wurtz2018cluster}, spin-boson models \cite{orioli2017nonequilibrium,mannouch2020partially,mannouch2020partially2,runeson2020generalized,runeson2019spin,cotton2013symmetrical}, and non-adiabatic molecular dynamics\cite{domcke2004conical} where the Born-Oppenheimer approximation breaks down.  \cite{meyer1979classical,meyera1979classical,cotton2013symmetrical,cotton2013symmetrical2,cotton2015symmetrical}.

In essence, TWA approaches treat bosons in the same way as mapping approaches treat the nuclei degrees of freedom (DoFs), examples being the phonons in trapped-ion experiments and bosonic ultracold atoms for TWA, and the nuclei in chemical reaction and photo-chemical experiments for mapping approaches. In contrast, there are several choices for the spin DoF (the electron subsystem). Consider an electron subsystem with $N$ electronic states, $\ket{1}, \ket{2},\cdots,\ket{N}$. The symmetry group of the electron DoF is $SU(N)$. MMST mapping approaches and Schwinger boson cluster TWA (CTWA) \cite{wurtz2018cluster} map the electron DoF to a single excitation of $N$ coupled Schwinger bosons, $b_1,b_2,\cdots,b_N$, or equivalently $N$ coupled harmonic oscillators, $X_1,P_1,X_2,P_2,\cdots,X_N,P_N$. A severe problem for MMST mapping approaches in the non-adiabatic dynamics is the physical phase space leakage problem, i.e., Schwinger bosons can escape from the single excitation phase space under the classical dynamics. This problem is partially solved by introducing a zero-point energy (ZPE) parameter that modifies the interaction between electronic and nuclei DoFs \cite{meyer1979classical,meyera1979classical,stock1999flow,muller1999flow}, or by a projection back to the single excitation Schwinger bosons phase space \cite{sun1998semiclassical,cotton2013symmetrical,cotton2013symmetrical2,hsieh2013analysis}. Instead, SM approaches, TWA, and Operator CTWA sample the spin DoF in the natural phase space of the $SU(2)$ \cite{meyer1979classical,cotton2015symmetrical} or $SU(N)$  group \cite{runeson2019spin,runeson2020generalized,mannouch2020partially,mannouch2020partially2}. All of the above methods use continuous DoFs to describe the electron subsystem. Recently, however, a novel TWA-related method based on Wooters' discrete phase space \cite{wootters1987wigner,gibbons2004discrete} for spins, the discrete Truncated Wigner Approximation (DTWA) \cite{schachenmayer2015many}, has been proposed and successfully generalized to higher spin systems (GDTWA) \cite{zhu2019generalized}. DTWA can capture the revivals and entanglement dynamics in  quantum spin lattice systems up to an astoundingly long time. Motivated by trapped-ion experiments, it has also been shown that DTWA is applicable to spin-boson models under the rotating wave approximation \cite{orioli2017nonequilibrium}. 

The goal of this work is to extend the scope of GDTWA to chemical systems, including a detailed theoretical analysis and numerical benchmarks. Our theoretical analysis shows that the discrete phase space used in GDTWA is tailor-made to treat the discrete space of electronic states in molecules.
Additional modifications often required to improve the accuracy of the existing mapping approaches, including a ZPE parameter\cite{stock1999flow,muller1999flow}, the projection back to the physical phase space\cite{sun1998semiclassical,cotton2013symmetrical,cotton2013symmetrical2,hsieh2013analysis}, and the different treatment of identity and traceless operators\cite{saller2019identity,saller2019improved}, are unnecessary in GDTWA.
The discrete phase space itself implicitly solves these mentioned issues.
As our numerical results illustrate, GDTWA achieves an accuracy at least as good as existing state-of-the-art mapping approaches, and outperforms
them in some of the selected applications in this article. 

This work is organized as follows. In Sec.~II, we introduce the GDTWA, first in its original formulation. By rewriting it in a language similar to the formulation of mapping approaches in chemistry, we show how to implement the simulations of GDTWA practically. In Sec.~III, we compare the GDTWA in the rewritten form with existing fully linearized methods to illustrate how GDTWA accounts for an effective ZPE without ZPE parameters, and we show how GDTWA differs from the  partially linearized methods. In Sec.~IV, we benchmark the GDTWA using two Linear Vibronic Coupling (LVC) models featuring non-adiabatic dynamics at a conical intersection. 
Section~\ref{sec:conclusions} contains our conclusions, and several Appendices complement the main text.

\section{Theory}

We first give the original form of the GDTWA. We then derive an equivalent form in analogous form to traditional mapping methods and the Ehrenfest method. This pedagogical rewriting allows us not only to implement the simulations with a lower computational cost; as further discussed in Sec. III, it also permits us to reveal special advantages of GDTWA, including the effective non-zero reduced ZPE and the absence of physical space leakage.

\subsection{Basics of GDTWA}

Consider a non-adiabatic Hamiltonian $\hat{H}$ describing $N$ electronic states, $\ket{1}, \ket{2},\cdots,\ket{N}$, coupled to a nuclear DoF (the generalization to several nuclear DoFs is straightforward). In the diabatic representation, we can write 
\begin{align}
    \label{eq:Ham}
    \hat{H} & = \frac{\hat{p}^2}{2m} + \hat{V}(\hat{x})\\\nonumber
            & = \frac{\hat{p}^2}{2m} + \sum_{kl}^{N}|k\rangle V_{kl}(\hat{x})\langle l|\,,
\end{align}
where $m$ is the mass of the nuclei, $\hat{x}$ and $\hat{p}$ are the nuclear coordinate and momentum operators. In this paper, we focus on initial product states of the form $\rho(0) = \rho_{\rm nuc}(0)\bigotimes\rho_{\rm el}(0)$. 
These can appear, e.g., in molecular systems with only one populated electronic state, such as the ground electronic state, or 
electronically excited systems prepared by a laser pulse shorter than the time-scale for nuclear displacements.

The density matrix of the electronic DoFs and the nuclei-electron interaction $\hat{V}(\hat{x})$ are matrices with $\mathcal{D} = N\times N$ elements. We can define $\mathcal{D}$ Hermitian operators $\hat{\Lambda}_\mu$, using the Generalized Gell-Mann Matrices (GGM) for $SU(N)$ \cite{bertlmann2008bloch} and the identity matrix $\rm \hat{I}$ as a complete basis for the electron DoF,
\begin{widetext}
\begin{equation}\label{eq:GGM}
    \hat{\Lambda}_\mu = \left\{
    \begin{aligned}
&\frac{1}{\sqrt{2}}(\ket{k}\bra{l} + \ket{l}\bra{k}) \quad \rm for \quad 1\le \mu \le N(N-1)/2, \quad 1\le l < k \le N\,,\\
&\frac{1}{\sqrt{2}i}(\ket{l}\bra{k} - \ket{k}\bra{l}) \quad \rm for \quad N(N-1)/2 < \mu \le N(N-1), \quad 1\le l < k \le N\,,\\
&\frac{1}{\sqrt{k(k+1)}}\sum_{l=1}^{k}(\ket{l}\bra{l} - k\ket{k+1}\bra{k+1}) \quad \rm for \quad N(N-1) < \mu \le N^2 - 1, \quad 1\le k < N\,,\\
&\sqrt{\frac{1}{N}}\hat{I} \rm \quad for \quad \mu = \mathcal{D}\,.
\end{aligned}
\right.
\end{equation}
\end{widetext}
The explicit form of the $\hat{\Lambda}_\mu$ for $N=2$ and $N=3$ are listed in the appendix~\ref{sec:ExplicitLambdamu}. 
The basis elements are orthonormal, $\trace{\hat{\Lambda}_\mu\hat{\Lambda}_\nu = \delta_{\mu\nu}}$ with the commutation relation $[\hat{\Lambda}_\mu,\hat{\Lambda}_\nu] = if_{\mu\nu\xi}\hat{\Lambda}_\xi$, where $f_{\mu\nu\xi}$ are the structure constants,
\begin{equation}
    if_{\mu\nu\xi} = \trace(\hat{\Lambda}_\xi[\hat{\Lambda}_\mu,\hat{\Lambda}_\nu])\,,
\end{equation}
and the Einstein notation has been used. 
We are going to use these basis elements to derive a semiclassical description. 

Any operator $\hat{O}_{\rm el}$ acting on the electron DoF can be expanded as $\sum_\mu c_\mu\hat{\Lambda}_\mu$ with $c_\mu = \trace{\hat{O}_{\rm el}\hat{\Lambda}_\mu}$. Then, the Hamiltonian in Eq.~(\ref{eq:Ham}) can be expressed as
\begin{equation}
\hat{H} = \frac{\hat{p}^2}{2m}\sqrt{N}\hat{\Lambda}_\mathcal{D} + \sum_\mu v_\mu(\hat{x})\hat{\Lambda}_\mu\,,
\end{equation}
with $v_\mu(\hat{x}) = \trace{\hat{V}(\hat{x})\hat{\Lambda}_\mu}$. The Heisenberg equation of motions (EOMs) of the operators are 
\begin{equation}\label{eq:EOMQ}
\begin{aligned}
    \dot{\hat{x}}_t & = \hat{p}_t/m\,, \\
    \dot{\hat{p}}_t & = -\partial_{{\hat{x}}_t}v_\mu(\hat{x}_t)\hat{\Lambda}_\mu(t)\,,  \\
    \dot{\hat{\Lambda}}_\mu(t) & = f_{\mu\nu\xi}v_\nu(\hat{x}_t)\hat{\Lambda}_\xi(t)\,.
\end{aligned}
\end{equation}

As in the usual linearized semiclassical methods, GDTWA approximates the observables as statistical averages over trajectories of the phase space variables whose equations of motion are classical and formally identical to the quantum Heisenberg EOMs. Define $x_t$, $p_t$, and $\lambda_\mu(t)$ as the time dependent classical phase variables for $\hat{x}$, $\hat{p}$, and $\hat{\Lambda}_\mu$, respectively. Then, their EOMs are
\begin{equation}\label{eq:EOMG}
\begin{aligned}
    \dot{x}_t & = p_t/m\,, \\
    \dot{p}_t & = -\partial_{x_t}v_\mu(x_t)\lambda_\mu(t)\,,  \\
    \dot{\lambda}_\mu(t) & = f_{\mu\nu\xi}v_\nu(x_t)\lambda_\xi(t)\,,
\end{aligned}
\end{equation}
~\\
with initial condition $x_{t=0}=x_0$ and $p_{t=0}=p_0$.
At this stage, the correlators between nuclei and electrons are taken classical, which amounts to taking the mean-field form of the Heisenberg EOMs in each single trajectory. That approach effectively truncates the order of the EOMs. Though the EOMs of GDTWA in each single trajectory are formally identical to the mean-field method, GDTWA is still a method beyond the mean-field theory because the quantum fluctuations are partially accounted for in the initial statistical distributions of the phase space variables, which is similar to traditional TWA and mapping approaches \cite{hillery1984distribution,steel1998dynamical,blakie2008dynamics,polkovnikov2010phase,schachenmayer2015many,zhu2019generalized,davidson2015s,wurtz2018cluster,polkovnikov2003quantum,orioli2017nonequilibrium,pucci2016simulation,meyera1979classical,cotton2013symmetrical,stock1997semiclassical,cotton2013symmetrical2,liu2017isomorphism,he2019new,liu2016unified,miller2017classical,saller2019identity,saller2019improved,sun1998semiclassical,kim2008quantum,kelly2012mapping,huo2011communication,huo2013communication,hsieh2012nonadiabatic,hsieh2013analysis,kapral1999mixed,stock1999flow,muller1999flow,huo2012consistent,cotton2015symmetrical,meyer1979classical,mannouch2020partially,mannouch2020partially2,runeson2019spin,runeson2020generalized}.

The sampling of GDTWA for the initial nuclear phase variables are identical to the ordinary linearized semiclassical methods, 
\begin{equation}
    W_{\rm nuc}(x_0,p_0) = \int d\eta \bra{x_0-\frac{\eta}{2}}\rho_{\rm nuc}(0)\ket{x_0+\frac{\eta}{2}}e^{ip_0\eta}\,.
\end{equation}

The novelty of GDTWA is to sample the initial $\lambda_\mu$ as a discrete distribution. The details are as follows. First, $\hat{\Lambda}_\mu$ can be decomposed as $\hat{\Lambda}_\mu = \sum_{a_\mu} a_\mu \ket{a_\mu}\bra{a_\mu}$, where $\ket{a_\mu}$ are the eigenvectors of $\hat{\Lambda}_\mu$. Then, the initial distribution of $\lambda_\mu(0)$ is $\lambda_\mu(0) \in \{a_\mu\}$ with probabilities
\begin{equation}\label{GSample}
 p(\lambda_\mu(0) = a_\mu) = \trace[\hat{\rho}_{el}(0)\ket{a_\mu}\bra{a_\mu}]\,.
\end{equation}
This distribution can represent arbitrary quantum expectation values exactly as a statistical average,
\begin{equation}
    \braket{\hat{O}_{\rm el}} = \sum_\mu c_\mu\braket{\hat{\Lambda}_\mu} = \sum_{\mu,a_\mu} c_\mu p(\lambda_\mu(0) = a_\mu)a_\mu\,.
\end{equation}

We are now in a position to give the formula to evaluate arbitrary observables $\hat{O} = \hat{O}_{\rm nuc}\bigotimes \hat{O}_{\rm el}$ under the GDTWA framework, 
\begin{widetext}
\begin{equation}\label{eq:ObsG}
\begin{aligned}
    \braket{\hat{O}(t)} & \approx \sum_{\mu,a_\mu}\int dx_0dp_0 W_{\rm nuc}(x_0,p_0)O_{w,\rm nuc}(x_t,p_t)  c_\mu p(\lambda_\mu(0) = a_\mu)\lambda_\mu(t)\,,
\end{aligned}
\end{equation}
\end{widetext}
where $O_{w,\rm nuc}$ is the Wigner transformation of the operator $\hat{O}_{\rm nuc}$
\begin{equation}
    O_{w,\rm nuc}(x,p) = \int d\eta \bra{x-\frac{\eta}{2}}\hat{O}_{\rm nuc}\ket{x+\frac{\eta}{2}}e^{ip\eta}.
\end{equation}

In principle, the above sampling can be applied to arbitrary electronic initial states. However, some specific initial electronic states result in a higher accuracy than others. Namely, an increased accuracy is achieved for initial states for which the statistical sampling reproduces the initial intra-correlation \cite{zhu2019generalized} of the electron states, i.e., for the observables
\begin{widetext}
\begin{equation}\label{eq:Intra}
    \braket{\frac{\hat{\Lambda}_\mu\hat{\Lambda}_\nu + \hat{\Lambda}_\nu\hat{\Lambda}_\mu}{2}} 
    =\sum_{a_\mu,a_\nu}
    p(\lambda_\mu(0)=a_\mu)p(\lambda_\nu(0)=a_\nu)a_\mu a_\nu \quad {\rm for}\quad \mu\neq \nu,
    \quad \braket{\hat{\Lambda}_\mu^2} =\sum_{a_\mu} p(\lambda_\mu(0)=a_\mu)a_\mu^2\,.
\end{equation}
\end{widetext}
A detailed analysis of the sampling of initial conditions can be found in the Appendix B.

Generally, it has been proven that the GDTWA sampling distribution can reproduce the intra-electron correlation for the diagonal states \cite{zhu2019generalized} $\ket{m}\bra{m}$, $1\le m\le N$.  For convenience, we only consider the initial state $\ket{1}\bra{1}$ in this article. All the other initial pure states can be converted to this state by unitary transformations, and all expectation values of observables of mixed states can be expressed as the summation over the expectation value of pure states. 

\subsection{Re-formulation of GDTWA in the language of mapping approaches }


In the following, we re-write the GDTWA in a completely equivalent form that not only reduces the computational cost by reducing the classical DoFs used to describe the electronic subsystem from $N^2-1$ to $4N$ \cite{wurtz2018cluster}, but also reveals important concepts such as ZPE (see Sec.~\ref{sec:ZPEinGDTWA}), thus enabling a direct comparison to the formalism of linearized semiclassical methods (see Sec.~\ref{sec:ZPEinGDTWA} and Sec.~\ref{sec:comparison_with_partially_linearized_methods}).

At the core of GDTWA lies a sampling over trajectories. In the original formulation of GDTWA, this is achieved via sampling over the continuous initial phase space of the nuclear degree of freedom as well as the discrete electronic initial phase space variables $\lambda^{(\alpha)}_\mu(0)$, where we used the index $\alpha$ to label the diverse electronic initial conditions in the discrete phase space. 
In the formulation we are developing here, the role of $\lambda^{(\alpha)}_\mu(0)$ is assumed by the so-called discrete quasi-phase point operators $A_\alpha(0)$, which are used to describe the electronic DoFs using the transformation 
\begin{equation}
\label{eq:lambdamu_Aalpha}
\begin{aligned}
    A_\alpha(t)=\sum_\mu \lambda^{(\alpha)}_\mu(t)\hat{\Lambda}_\mu\\
    \lambda^{(\alpha)}_\mu(t) = \trace{A_\alpha(t)\hat{\Lambda}_\mu}\,.
\end{aligned}
\end{equation}
For convenience, we will use the notation $A_\alpha$ to express $A_\alpha(t)$ in this article when there is no ambiguity.

The sampling of the initial condition $A_\alpha(0)$ is achieved via a sampling of the initial $\lambda^{(\alpha)}(0)$ as in Eq.~\eqref{GSample}, which using the transformation Eq.~\eqref{eq:lambdamu_Aalpha} translates into
\begin{equation}\label{Aalpha}
    A_\alpha(0) = \left( \begin{array}{cccc}
1 & \frac{\delta_2 - i\sigma_2}{2} & \cdots & \frac{\delta_N - i\sigma_N}{2}\\
\frac{\delta_2 + i\sigma_2}{2} & 0 & \cdots & 0\\
\vdots & \vdots & \ddots & \vdots \\
\frac{\delta_N + i\sigma_N}{2} & 0 & \cdots & 0\\
\end{array} 
\right ),
\end{equation}
with $\delta_i, \sigma_i = \pm 1$ being independent and identically distributed discrete uniform variables on the integers $\pm 1$. The initial density matrix of the electron subsystem is expanded as $\rho_{\rm el}(0) = \ket{1}\bra{1} = \sum_\alpha w_\alpha A_\alpha(0)$, where $w_\alpha = 2^{-2(N-1)} $ for all $\alpha$. The GDTWA sampling strategy for the electron subsystem is converted to generating the initial discrete phase points by sampling $\delta_i$ and $\sigma_i$ accordingly. In fact, $A_\alpha(0)$ is nothing but the quasi-phase point operator in the Wootters' discrete phase space representation \cite{wootters1987wigner,gibbons2004discrete,note}.

The ansatz of GDTWA in this form is that the Wigner function is evolved along the classical stationary trajectories
\begin{equation}
\begin{aligned}
    W(x,p,A,t) \approx \sum_\alpha \int dx_0dp_0 w_\alpha W_{\rm nuc}(x_0,p_0)\\
    \delta(x-x_t)\delta(p-p_t) \bigotimes A_\alpha(t) ,
\end{aligned}
\end{equation}
where the EOMs of the variables are
\begin{equation}\label{eq:EOM}
\begin{aligned}
    \dot{x}_t & = p_t/m\,, \\
    \dot{p}_t & = -\partial_{x_t} \Tr{A_\alpha(t)\hat{V}(x_t)}\,, \\
    \dot{A}_\alpha(t) & = i[A_\alpha(t),\hat{V}(x_t)]\,,
\end{aligned}
\end{equation}
with initial condition $x_{t=0}=x_0$ and $p_{t=0}=p_0$. Any observable $\hat{O} = \hat{O}_{\rm nuc}\bigotimes \hat{O}_{\rm el}$ can be evaluated as 
\begin{equation}\label{eq:Obs}
\begin{aligned}
    \braket{\hat{O}(t)} & \approx \trace\int dxdp W(x,p,A,t)O_{w,\rm nuc}(x,p)\bigotimes\hat{O}_{\rm el} \\
& = \sum_\alpha \int dx_0dp_0 w_\alpha W_{\rm nuc}(x_0,p_0)O_{w,\rm nuc}(x_t,p_t)\\
&\phantom{=}\times \Tr{A_\alpha(t)\hat{O}_{\rm el}}\,.
\end{aligned}
\end{equation}

The GDTWA in this form, with the EOMs given by Eq.~(\ref{eq:EOM}) and the expectation values in Eq.~(\ref{eq:Obs}),
has some formal resemblances to the Ehrenfest method. In both approaches, each trajectory of the nuclei evolves in the mean potential
resulting from the populated electronic states. However, there are two main differences between these two methods. First, GDTWA trajectories start from a discrete sampling in the space of the quasi-phase point operators
rather than from a uniquely defined electron state.
Second, GDTWA trajectories evolve the quasi-phase point operator $A_\alpha(t)$ rather than $\rho_{\mathrm{el}}(t)$ in each individual trajectory.

To implement the simulation, we require the spectral decomposition for the quasi-phase point operator $A_\alpha$. It is easy to check that the spectral decomposition of Eq.~(\ref{Aalpha}) is $A_\alpha(0) = \lambda_+\ket{\Psi_+^\alpha(0)}\bra{\Psi_+^\alpha(0)} + \lambda_-\ket{\Psi_-^\alpha(0)}\bra{\Psi_-^\alpha(0)}$, where the eigenvalues are
\begin{equation}
    \lambda\pm = \frac{1\pm \sqrt{2N-1}}{2}\,,
\end{equation}
with the amplitudes of the associated eigenvectors
\begin{equation}
\begin{aligned}
    \braket{1|\Psi_{\pm}^\alpha(0)} & = \sqrt{\frac{\lambda_{\pm}^2}{\lambda_{\pm}^2 + (N-1)/2}}\,,\\
    \braket{j|\Psi_{\pm}^\alpha(0)} & = \sqrt{\frac{\lambda_{\pm}^2}{\lambda_{\pm}^2 + (N-1)/2}}\frac{\delta_j + i\sigma_j}{2\lambda_{\pm}}\qquad \forall j > 1\,.
\end{aligned}
\end{equation}
The eigenvalues of the quasi-phase point operator can be interpreted as quasi-probabilities, since $\lambda_+ + \lambda_- = 1$, $\lambda_+ > 0$ and $\lambda_- < 0$. Such quasi-probabilities constitute the spectrum of $A_\alpha$, and are conserved during the propagation. We can propagate $\ket{\Psi_\pm^\alpha (t)}$ rather than $A_\alpha(t)$ using the EOMs 
\begin{equation}
    i\frac{d}{dt}\ket{\Psi_\pm^\alpha (t)}=\hat{V}(x_t)\ket{\Psi_\pm^\alpha (t)}
\end{equation}
and $A_\alpha(t)=\lambda_+\ket{\Psi_+^\alpha(t)}\bra{\Psi_+^\alpha(t)} + \lambda_-\ket{\Psi_-^\alpha(t)}\bra{\Psi_-^\alpha(t)}$.
This completely equivalent reformulation reduces the number of electronic subsystem DoFs
from $N^2-1$ to $4N$.

\section{Discussion}
In this section, we compare the GDTWA with established fully and partially linearized semiclassical methods. The form of the EOMs of GDTWA is similar to fully linearized methods but with a computational cost close to partially linearized methods. Readers who are only interested in the numerical performance of GDTWA may skip this section.

\subsection{Zero point energy treatment within the GDTWA approach \& absence of physical space leakage}
\label{sec:ZPEinGDTWA}

Because of the discrete sampling, GDTWA accounts for a non-zero effective reduced ZPE without introducing an explicit ZPE parameter.
It is well known that both full ZPE (approaches based on MMST mapping without empirical ZPE parameters)
and zero ZPE (Ehrenfest method) are harmful for numerical accuracy \cite{stock1999flow,muller1999flow}. One possible solution to this problem is to introduce an adjusted ZPE-parameter to make the classical dynamics and phase space of the mapping variables of the harmonic oscillators of the electronic DoFs mimic the spin as much as possible \cite{runeson2019spin,runeson2020generalized,stock1999flow,muller1999flow}.
GDTWA solves this problem with a fundamentally different logic, i.e., GDTWA never introduces such a parameter but tames the ZPE only through a judiciously designed initial sampling procedure.

To illustrate how GDTWA accounts for an effective non-zero reduced ZPE, it is convenient to first review how existing methods including symmetrical quasi-classical windowing\cite{cotton2013symmetrical,cotton2013symmetrical2} and generalized spin mapping\cite{runeson2019spin,runeson2020generalized}, account for the ZPE. The EOMs of fully linearized mapping approaches \cite{meyera1979classical,cotton2013symmetrical,stock1997semiclassical,cotton2013symmetrical2,liu2017isomorphism,he2019new,liu2016unified,miller2017classical,saller2019identity,saller2019improved,sun1998semiclassical,kim2008quantum,kelly2012mapping,kapral1999mixed,stock1999flow,muller1999flow,huo2012consistent,runeson2019spin,runeson2020generalized} can also be written in the form of Eq.~(\ref{eq:EOM}), 
\begin{equation}\label{eq:EOMB}
\begin{aligned}
    \dot{x}_t & = p_t/m\,, \\
    \dot{p}_t & = -\partial_{x_t} \Tr{B_\alpha(t)\hat{V}(x_t)}\,, \\
    \dot{B}_\alpha(t) & = i[B_\alpha(t),\hat{V}(x_t)]\,,
\end{aligned}
\end{equation}
where 
\begin{equation}
    \label{eq:Balpha}
    B_\alpha(t) = R_\alpha^2\ket{\Psi_\alpha(t)}\bra{\Psi_\alpha(t)} - \frac{\gamma}{2}\hat{I}\,,    
\end{equation}
with $\gamma$ the ZPE parameter, usually chosen from zero (zero ZPE treatment) to one (full ZPE treatment), and $\ket{\Psi_\alpha(t)}$ the normalized electronic wave function. Further, $R_\alpha$ is the square root of the radius of the mapping variables, which in the ordinary harmonic oscillator MMST mapping notation, with position $X_n$ and momentum $P_n$ for state $n$, is defined by
\begin{equation}
    X_n(t) + iP_n(t) = \sqrt{2}R_\alpha\braket{n|\Psi_\alpha(t)}\,,
\end{equation}
\begin{equation}
    \sum_n X_n(t)^2 + P_n(t)^2 = 2R_\alpha^2.
\end{equation}
$R_\alpha$ and $\gamma$ are conserved during the evolution and the EOM of $\ket{\Psi_\alpha(t)}$ is 
\begin{equation}
    i\frac{d}{dt}\ket{\Psi_\alpha (t)}=\hat{V}(x_t)\ket{\Psi_\alpha (t)}\,.
\end{equation}

Different mapping approaches use different sampling strategies for $R_\alpha$ and $\ket{\Psi_\alpha(0)}$ and evaluate the expectation values of the observables in different manners. For each single trajectory, $B_\alpha(t)$ has one non-degenerate eigenvalue $R_\alpha^2 - \gamma/2$ and $N-1$ degenerate eigenvalues $-\gamma/2$, as can be seen immediately from the definition of $B_\alpha(t)$ in Eq.~\eqref{eq:Balpha}. In this sense, the ZPE parameter in the traditional fully linearized method is a negative diagonal energy correction term for the nuclei-electron interactions. The nuclei always see a modified average potential energy during the evolution in each single trajectory, whence mapping approaches with a non-zero ZPE parameter already account for some quantum effects in their EOMs. 

Though Eq.~(\ref{eq:EOMB}) and Eq.~(\ref{eq:EOM}) are formally identical, it is impossible to express $A_\alpha$ in the form $R_\alpha^2\ket{\Psi_\alpha(t)}\bra{\Psi_\alpha(t)} - \frac{\gamma}{2}\hat{I}$, and thus to construct the ZPE-parameter, except for the case of $N=2$, in which case, $\gamma = \sqrt{3} - 1$. We can nevertheless identify an effective ZPE-parameter governing the evolution of $A_\alpha$. Namely, the ZPE-parameter in the traditional fully linearized methods can also be constructed by the following strategy. Notice that $\trace(B_\alpha) = R_\alpha^2 - \frac{\gamma}{2}N$ and $\trace(B_\alpha^2) = R_\alpha^4 - \gamma R_\alpha^2 + \frac{\gamma^2}{4}N$ only depend on $R_\alpha$ and $\gamma$. Hence, the ZPE-parameter in the traditional fully linearized methods can be expressed as 
\begin{equation}\label{eq:EffZPE}
\begin{aligned}
    \gamma = \frac{ \sqrt{N\trace{(B_\alpha^2)}-(\trace{B_\alpha})^2 }}{N\sqrt{N-1}}
    - \frac{\trace B_\alpha }{N}
\end{aligned}
\end{equation}
By formally replacing $A_\alpha$ with $B_\alpha$ in Eq.~(\ref{eq:EffZPE}), we obtain an effective ZPE-parameter for the GDTWA,
\begin{equation}\label{eq:EffZPEGDTWA}
\begin{aligned}
    \gamma_{\rm eff} = \frac{2\sqrt{N+1}-2}{N}\,.
\end{aligned}
\end{equation}
Interestingly, this reduced effective ZPE coincides with the ZPE in recent works using the SM approach \cite{runeson2019spin,runeson2020generalized,mannouch2020partially,mannouch2020partially2}. The reason of such identical ZPE is that both GDTWA and SM start from the phase space of the electronic DoFs, rather than the phase space of Schwinger bosons. The ZPE of SM and GDTWA can, however, be different when the Hamiltonian is block diagonal, see the discussions in the Appendix C. 

A further feature of the implicit ZPE treatment is that GDTWA treats the traceless and identity operators of electronic states in a unified way. No other trick \cite{saller2019identity,saller2019improved} or a specific implementation for the identity operator \cite{mannouch2020partially,mannouch2020partially2} is required. In this sense, GDTWA seems a more natural approach to obtain observables of the electronic DoF.

Another advantage related to the spin phase space of GDTWA is that the method does not suffer from the physical space leakage problem \cite{stock1999flow,muller1999flow}, and thus eliminates the additional projection that is necessary in the LSC-IVR and PBME approaches \cite{sun1998semiclassical,cotton2013symmetrical,cotton2013symmetrical2,hsieh2013analysis}. The EOMs and initial sampling constructions ensure that the $A_\alpha(t)$ trajectories are always trapped in this tailor-made electronic phase space, similarly to what is achieved for $B_\alpha(t)$ in the recently proposed SM approach\cite{runeson2019spin,runeson2020generalized,mannouch2020partially,mannouch2020partially2}. 

\subsection{Comparison with partially linearized methods}
\label{sec:comparison_with_partially_linearized_methods}

The nuclei in both GDTWA and partially linearized methods move on a mean-field potential, which is the average potential of two effective electronic states, in each single trajectory. Nevertheless, GDTWA has a significantly different logic from traditional partially linearized methods, such as the Forward-Backward Trajectory solution (FBTS) \cite{hsieh2012nonadiabatic,hsieh2013analysis}, partially Linear Density Matrix (PLDM) \cite{huo2011communication,huo2013communication}, and Spin-PLDM \cite{mannouch2020partially,mannouch2020partially2}, as we illustrate now.

The EOMs of the family of partially linearized methods can be written as \cite{hsieh2012nonadiabatic,hsieh2013analysis,huo2011communication,huo2013communication,mannouch2020partially,mannouch2020partially2}
\begin{equation}\label{eq:EOMFB}
\begin{aligned}
    \dot{x}_t & = p_t/m\,, \\
    \dot{p}_t & = -\frac{R_{1,\alpha}^2}{2}\partial_{x_t} \bra{\Psi_{1,\alpha}(t)}\hat{V}(x_t)\ket{\Psi_{1,\alpha}(t)}\\
    & -\frac{R_{2,\alpha}^2}{2}\partial_{x_t} \bra{\Psi_{2,\alpha}(t)}\hat{V}(x_t)\ket{\Psi_{2,\alpha}(t)}\,, \\
    i\frac{\rm d}{\rm dt}\ket{\Psi_{1,\alpha}(t)} & = \hat{V}(x_t)\ket{\Psi_{1,\alpha}(t)}\,, \\
    i\frac{\rm d}{\rm dt}\ket{\Psi_{2,\alpha}(t)} & = \hat{V}(x_t)\ket{\Psi_{2,\alpha}(t)}\,, 
\end{aligned}
\end{equation}
where $\ket{\Psi_{1,\alpha}(t)}$ and $\ket{\Psi_{2,\alpha}(t)}$ are the forward and backward normalized electronic wavefunctions (or electronic trajectories), respectively, and $R_{1,\alpha}$ and $R_{2,\alpha}$ are the square root of the radius of the corresponding mapping variables. In the ordinary harmonic oscillator MMST mapping notation,
\begin{equation}
    X_{j,n}(t) + iP_{j,n}(t) = \sqrt{2}R_{j,\alpha}\braket{n|\Psi_{j,\alpha}(t)}\,,
\end{equation}
\begin{equation}
    \sum_n X_{j,n}(t)^2 + P_{j,n}(t)^2 = 2R_{j,\alpha}^2,\quad {\rm for} \quad j  = 1,2.
\end{equation}

Different partially linearized methods have different formulas to evaluate expectation values and different sampling strategies for the initial radius and electronic trajectories. The electronic subsystem in each single trajectory of different partially linearized methods are also different. A typical electronic subsystem in partially linearized methods takes the form $\ket{\Psi_{1,\alpha}(t)}\bra{\Psi_{2,\alpha}(t)}$, which, unlike $A_\alpha(t)$ and $B_\alpha(t)$, is not hermitian. Specifically, the sampling of $\ket{\Psi_{1,\alpha}(0)}$ and $\ket{\Psi_{2,\alpha}(0)}$ must be uncorrelated. As a comparison, there is no forward and backward electronic trajectories concept in GDTWA. So, the two electronic wavefunctions for GDTWA are the spectral decomposition of the quasi-phase point operator. The initial conditions for two electronic states in GDTWA in a single trajectory are necessarily correlated. In this sense, GDTWA is a method with hybrid features of fully linearized methods and partially linearized methods, i.e., GDTWA has the same form of EOMs as the fully linearized methods, but two electronic wavefunctions in each single trajectory. In conjunction with the inclusion of an effective ZPE as well as two electronic states in each single trajectory, this makes GDTWA an extremely efficient and surprisingly reliable numerical method, as we will see in the numerical computations of the following section.

\section{Numerical Results}
\label{sec:numerical_results}

In this section, we perform numerical benchmarks on the GDTWA for prototypical non-adiabbatic dynamics problems in chemistry.
Since each GDTWA trajectory evolves the classical nuclei and
two coupled electronic time-dependent states, its numerical complexity is close to the partially linearized approach and slightly larger than the fully linearized mapping approach. We may thus expect that GDTWA should be considered as an alternative approach to partially linearized methods, which is indeed confirmed by the numerics reported in this section. The selected mapping approaches to which we compare in this section are  PLDM\cite{huo2011communication}, Spin-PLDM\cite{mannouch2020partially,mannouch2020partially2} with non-focus sampling, and the Ehrenfest\cite{domcke2004conical}
method. For all the methods we run $10^6$ trajectories to ensure convergence, though GDTWA starts to converge already with $10^4$ trajectories, a number comparable with the Ehrenfest method. We will show numerical benchmarks for two LVC models \cite{domcke2004conical,koppel1988interplay,koppel1993new,schneider1989surface}, comparing the selected linearized semiclasscial methods with numerically converged Multi-configuration time-dependent Hartree (MCTDH) calculations \cite{manthe1992wave,meyer1990multi,beck2000multiconfiguration}. 

The LVC Hamiltonian \cite{koppel1984multimode,domcke1997theory}
in the diabatic basis is given by
\begin{equation}
    H = \frac{1}{2}\sum_j \omega_j p_j^2 + \sum_{k,l}\ket{k}W_{kl}\bra{l}\,,
\end{equation}
where $W_{kl}$ is obtained by the Taylor expansion with respect to the electronic ground state
equilibrium geometry, 
\begin{align}
    W_{kk} &= E_k + \frac{1}{2}\sum_j \omega_jx_j^2 + \sum_j \kappa_j^{(k)}x_j\,,\\
    W_{kl} &= \sum_j \lambda_j^{(kl)}x_j, \quad k\neq l\,,
\end{align}
where $x_j$ and $p_j$ are the dimensionless position and momentum for the vibronic mode $j$, and $\omega_j$ is the corresponding frequency. Further, $E_k$ is the vertical transition energy of the diabatic state $\ket{k}$, and $\lambda_j^{(kl)}$ and $\kappa_j^{(k)}$ are the gradients of $W_{kl}$ and $W_{kk}$, respectively.

In this article, we focus on the time dependence of observables for the initial product state of the vibrational ground state $\Psi = \prod_j \frac{1}{\pi^{1/4}}\exp{-x_j^2/2}$ and the excited electronic state, which is a typical setup of femtochemistry experiments. 
We consider two typical benchmark models \cite{domcke2004conical,koppel1988interplay,koppel1993new,schneider1989surface}, as given in the Tables~\ref{tab:ModelI} and~\ref{tab:ModelII}. Model I is a three-modes two-states model based on Pyrazine. It includes two tuning coordinates $x_1$ and $x_{6a}$, and one coupling coordinate $x_{10a}$, and the initial electron wave function is prepared in the second diabatic state $\ket{2}$ \cite{domcke2004conical}. Model II is a five-modes three-states model based on Benzene radical cation. It includes three tuning coordinates $x_2$, $x_{16}$, and $x_{18}$, and two coupling coordinates $x_{8}$ and $x_{19}$, and the electron wave function is initialized in the third diabatic state $\ket{3}$ \cite{domcke2004conical}.

\begin{table*}
\label{Table.1}
\begin{center}
\begin{tabular*} {15 cm} {@{\extracolsep{\fill} }cccccccc}
\toprule
\hline
\quad & $E_k$ & $\omega_1$ & $\kappa_1^{(k)}$ & $\omega_{6a}$ & $\kappa_{6a}^{(k)}$ & $\omega_{10a}$ & $\lambda$ \\

\hline
$\ket{1}$ & 3.94 & 0.126 & 0.037 & 0.074 & −0.105 & 0.118\\
\quad &\quad &\quad &\quad &\quad &\quad &\quad & 0.262\\
$\ket{2}$ & 4.84 & 0.126 & −0.254 & 0.074 & 0.149 & 0.118\\

\hline
\end{tabular*}
\end{center}
\caption{Parameters of Model I that is based on Pyrazine. All quantities are given in eV.}
\label{tab:ModelI}
\end{table*}

\begin{table*}
\label{Table.2}
\begin{center}
\begin{tabular*} {15 cm} {@{\extracolsep{\fill} }cccccccccccc}
\toprule
\hline
\quad & $E_k$ & $\omega_2$ & $\kappa_2^{(k)}$ & $\omega_{16}$ & $\kappa_{16}^{(k)}$ & $\omega_{18}$ & $\kappa_{18}^{(k)}$ & $\omega_{8}$ & $\lambda_8^{(12)}$ & $\omega_{19}$ & $\lambda_{19}^{(23)}$\\

\hline
$\ket{1}$ & 9.75 & 0.123 & -0.042 & 0.198 & -0.246 & 0.075 & -0.125 & 0.088 & \quad & 0.12 & \quad\\
\quad &\quad &\quad &\quad &\quad &\quad &\quad &\quad &\quad &0.164\\
$\ket{2}$ & 11.84 & 0.123 & -0.042 & 0.198 & 0.242 & 0.075 & 0.1 & 0.088 & \quad & 0.12 & \quad\\
\quad &\quad &\quad &\quad &\quad &\quad &\quad &\quad &\quad &\quad &\quad &0.154\\
$\ket{3}$ & 12.44 & 0.123 & -0.301 & 0.198 & 0 & 0.075 & 0 & 0.088 & \quad & 0.12 & \quad\\

\hline
\end{tabular*}
\end{center}
\caption{Parameters of Model II based on Benzene radical cation. All quantities are given in eV.}
\label{tab:ModelII}
\end{table*}

\begin{figure}
	\centerline
	{\includegraphics[width=3.37in]{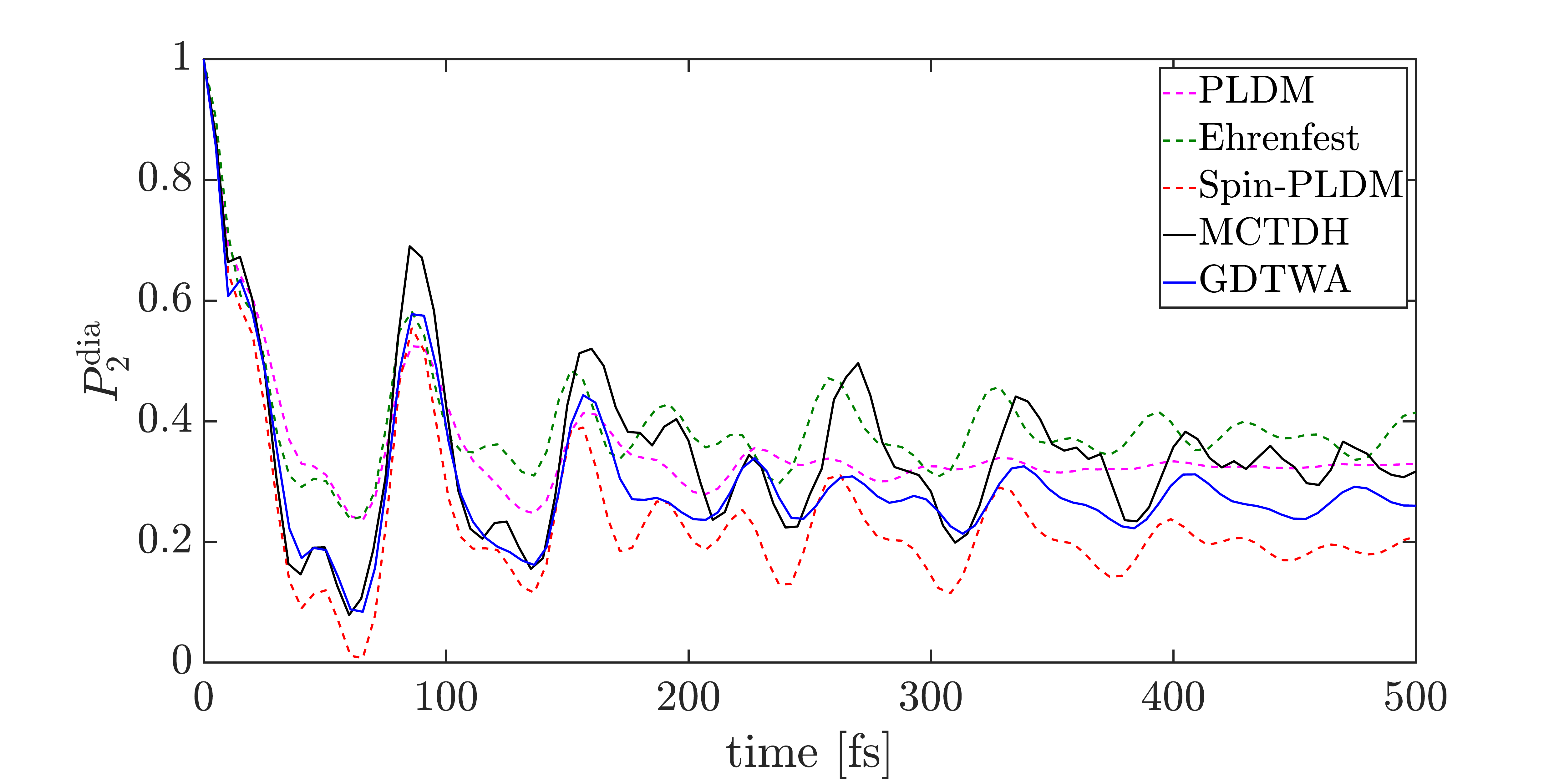}}
	\caption{Second diabatic state population of a three-modes two-states model based on Pyrazine (see table \ref{tab:ModelI}), computed using different methods. The GDTWA result (blue solid line) compares fairly well to the exact quantum dynamics (black solid). While GDTWA and, even more so, the Spin-PLDM method (red dashed) underestimate the mean value reached at long times, the Ehrenfest method (green dashed) overestimates it. The PLDM methods (pink dashed) considerably overestimates the damping of the oscillations. 
	}
	\label{fig:ThreeTwo}
\end{figure}

\begin{figure}
\centerline
{\includegraphics[width=3.37in]{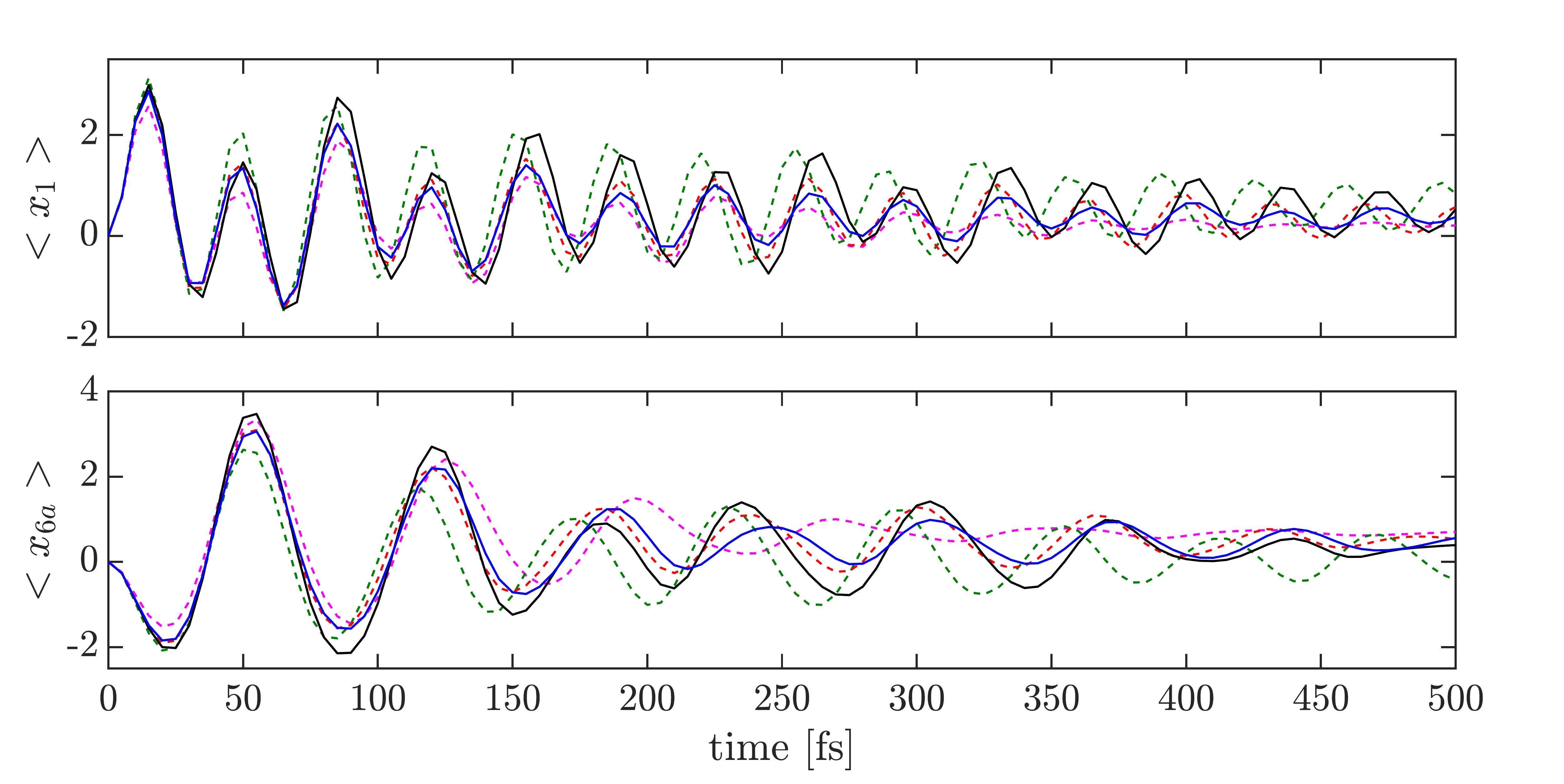}}

\caption{Populations of the tuning coordinates $\braket{x_1}$ and $\braket{x_{6a}}$ of the Pyrazine-based Model I. The color notations are identical to Fig.~\ref{fig:ThreeTwo}. The GDTWA (blue solid line) and Spin-PLDM (red dashed) results fail to capture the oscillation amplitudes, but still give a qualitatively fair description on the frequency. In contrast, the Ehrenfest (green dashed) and PLDM methods (pink dashed) mismatch the oscillation pattern of the exact quantum results (black dashed) after a few periods.
}
\label{TT}
\end{figure}

\begin{figure}
\centerline{\includegraphics[width=3.37in]{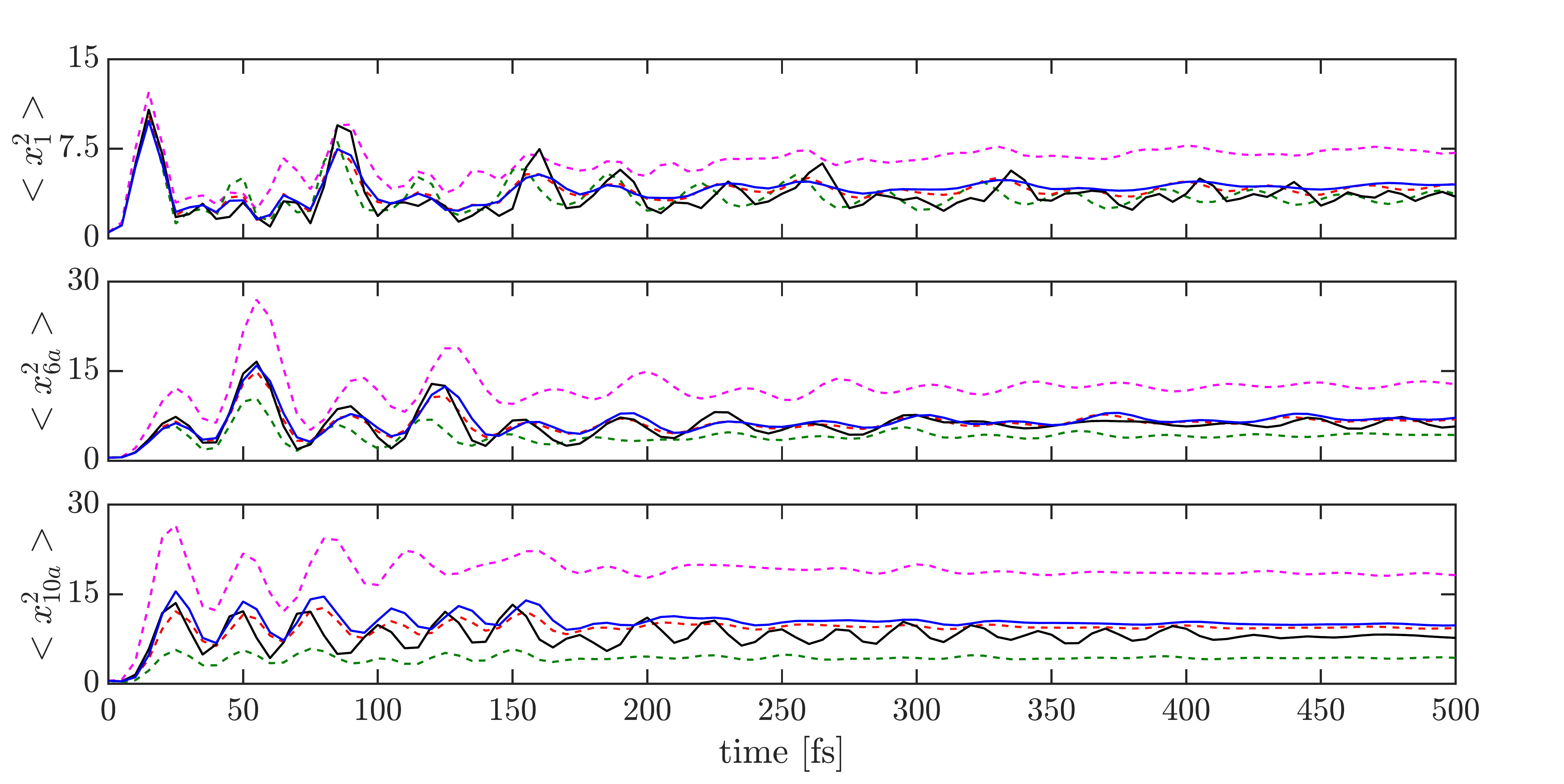}}

\caption{Expectation values of second-order correlations of the tuning coordinates $\braket{x_1^2}$ and $\braket{x_{6a}^2}$, and the coupling coordinate $\braket{x_{10a}^2}$ of the Pyrazine-based Model I. The color notations are identical to Fig.~\ref{fig:ThreeTwo}. The GdTWA (blue solid line) and Spin-PLDM (red dashed) results qualitatively predict the ideal higher-order correlation, while the Ehrenfest (green dashed) and PLDM methods (pink dashed) deviate significantly from the exact quantum results (black dashed).
}
\label{TC2}
\end{figure}

\begin{figure}
\centerline{\includegraphics[width=3.37in]{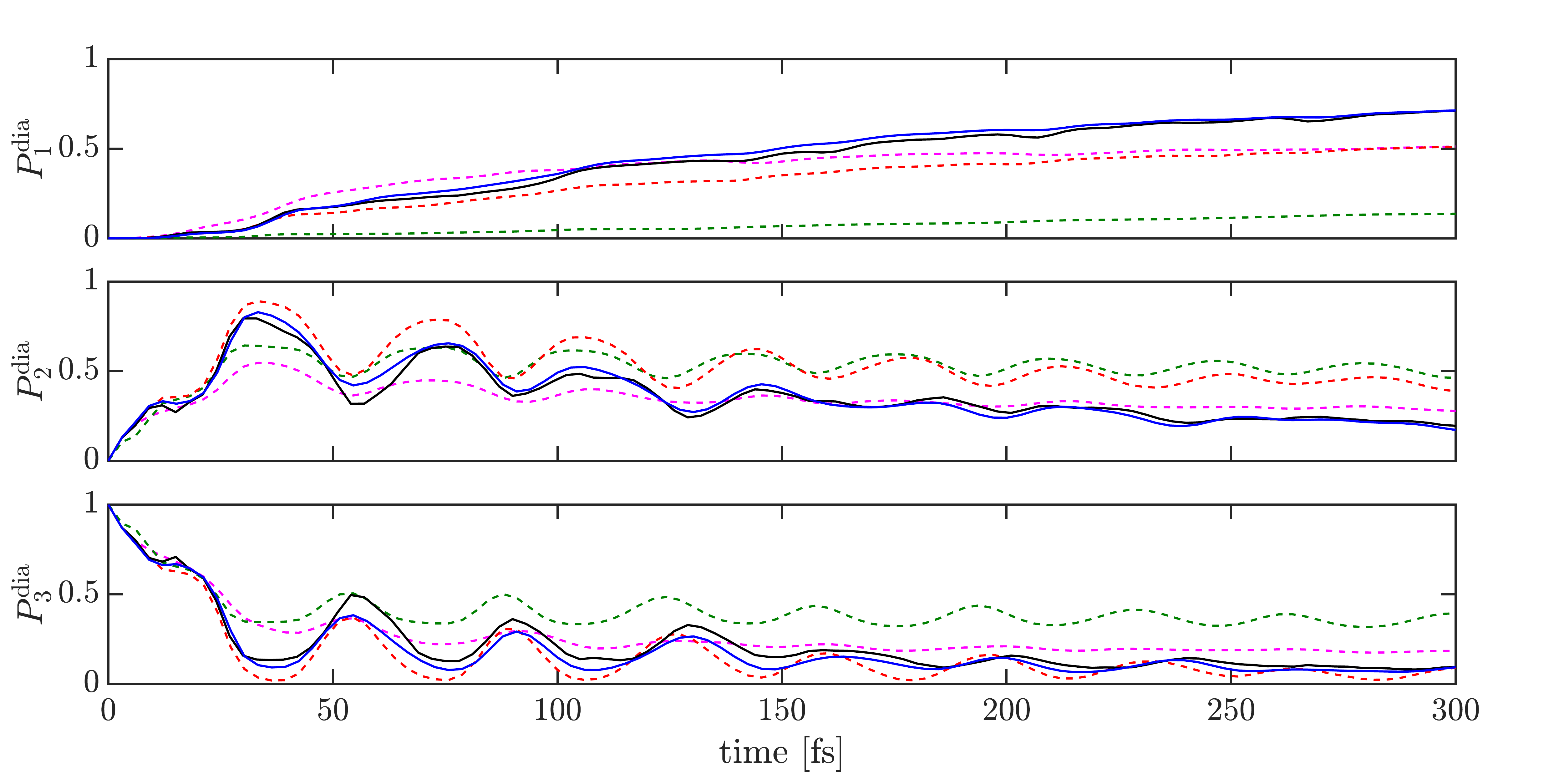}}
	\caption{
	Populations of all three diabatic states of a five-modes three-states model based on Benzene radical cation (see table \ref{tab:ModelII}), computed using different semiclassical techniques. The GdTWA result (blue solid line) compares fairly well to the exact quantum dynamics (black solid) for all the three diabatic states populations, while all the other methods considered fail to describe the long time populations.
	}
\label{FiveThree}
\end{figure}

\begin{figure}
\centerline{\includegraphics[width=3.37in]{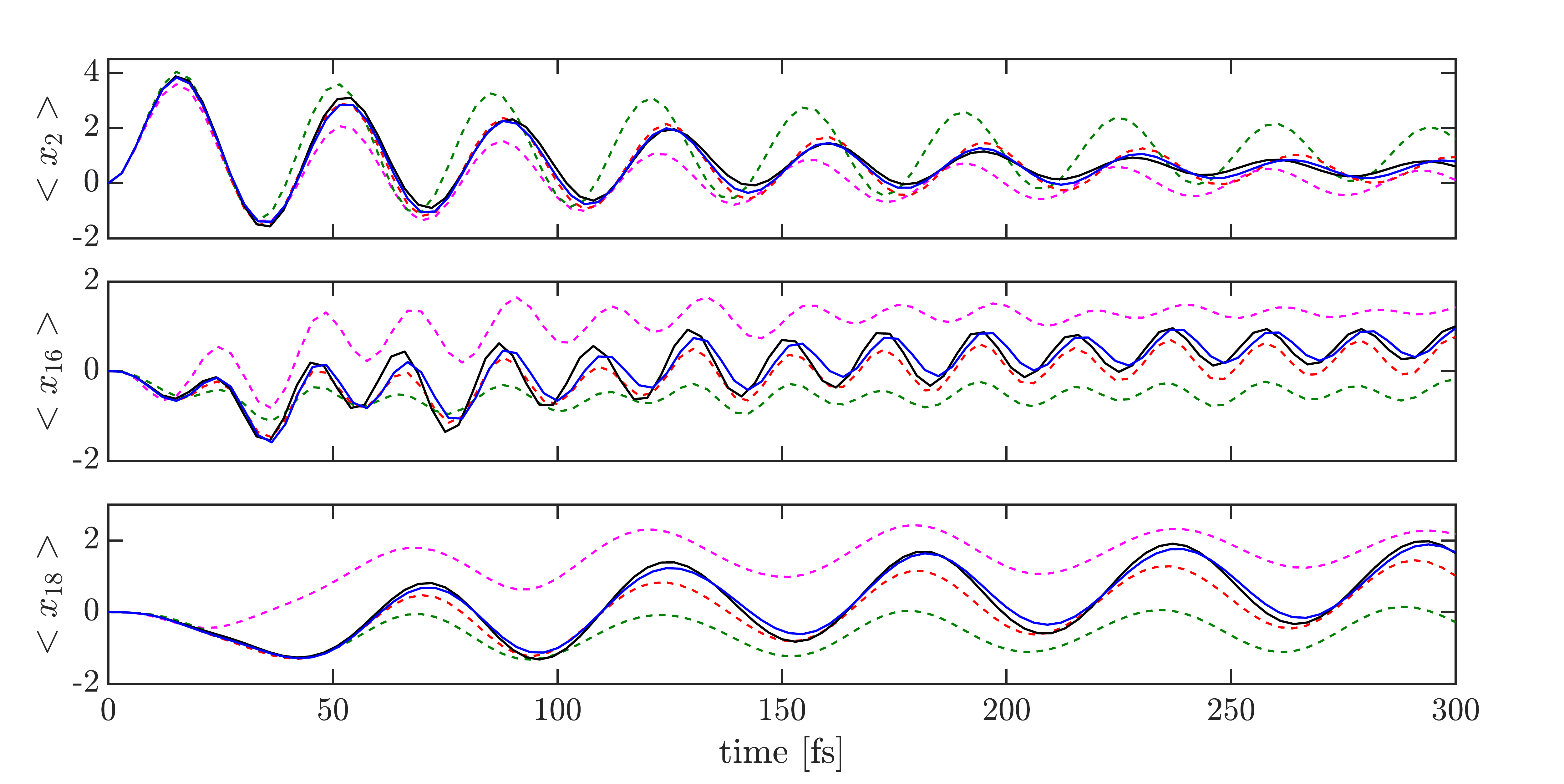}}

	\caption{
	Populations of tuning coordinates $\braket{x_2}$, $\braket{x_{16}}$, and $\braket{x_{18}}$ of the Model II that is based on Benzene radical cation. The GdTWA result (blue solid line) matches the exact quantum dynamics (black solid) best and slightly outperforms the Spin-PLDM result (red dashed).
	}
\label{FT}
\end{figure}

\begin{figure}
\centerline{\includegraphics[width=3.37in]{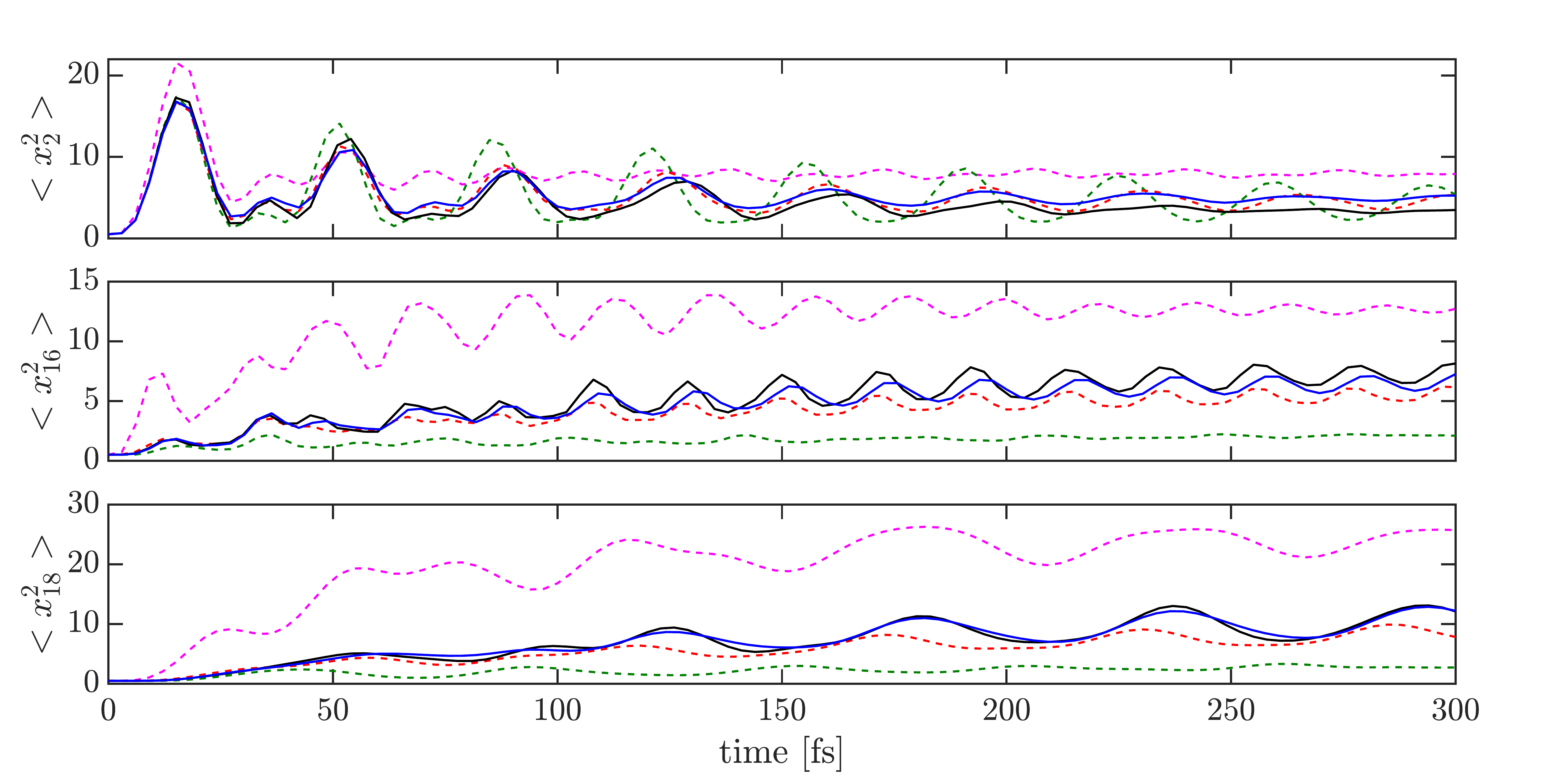}}
	\caption{
	The second-order correlations of the tuning coordinates $\braket{x_2^2}$, $\braket{x_{16}^2}$, and $\braket{x_{18}^2}$ of Model II. Both GDTWA (blue solid line) and Spin-PLDM (red dashed) match the exact quantum results (black solid) for the dynamics of $\braket{x_2^2}$. GDTWA slightly outperforms the Spin-PLDM result (red dashed) for the dynamics of $\braket{x_{16}^2}$, while GDTWA is noticeably more accurate than all the other methods for the dynamics of $\braket{x_{18}^2}$.
	}
\label{FT2}
\end{figure}

\begin{figure}
\centerline{\includegraphics[width=3.37in]{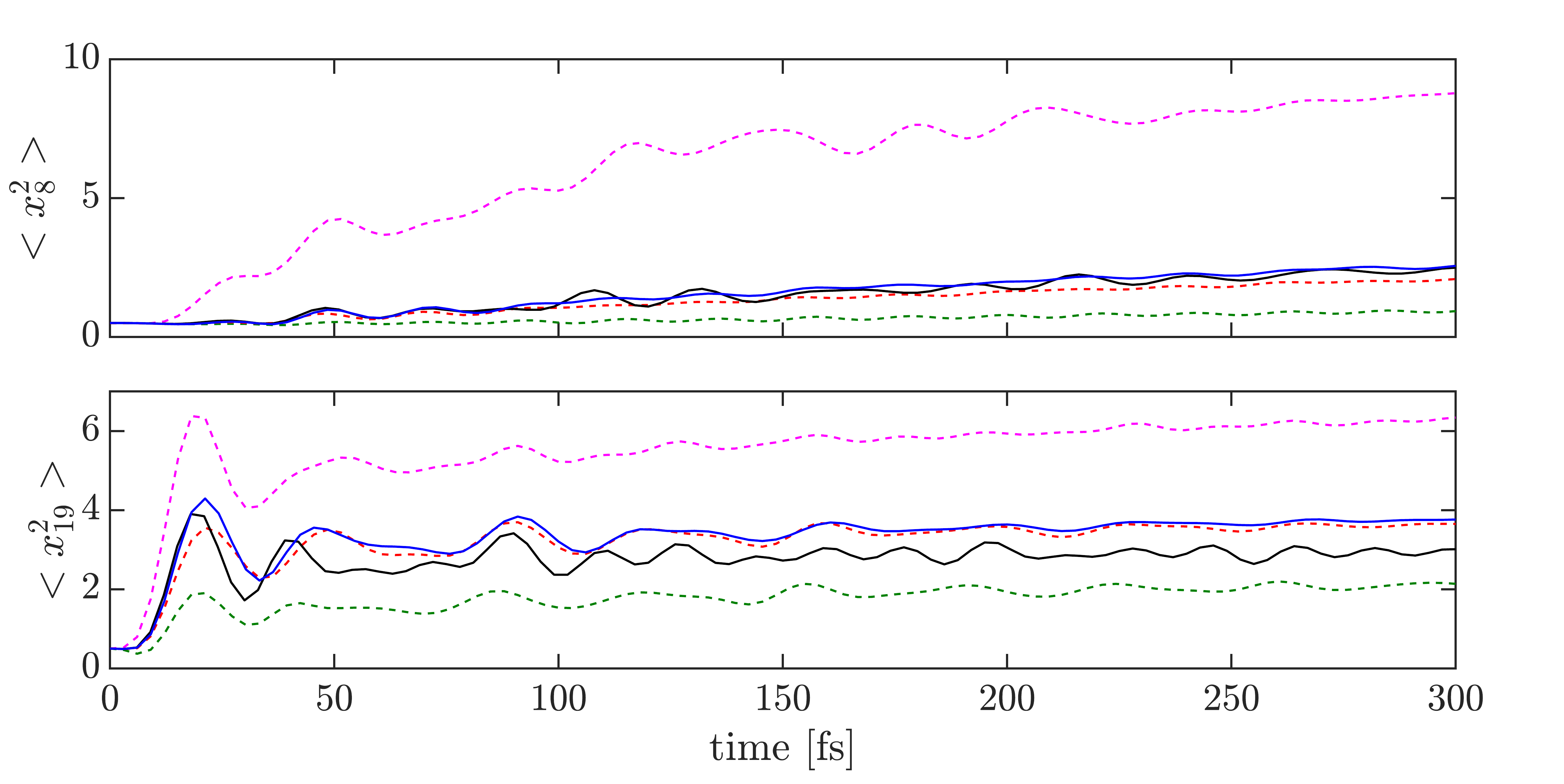}}
	\caption{
	Second-order correlations of the coupling coordinates $\braket{x_8^2}$ and $\braket{x_{19}^2}$ of Model II. For the dynamics of $\braket{x_8^2}$, both GDTWA (blue solid line) and Spin-PLDM (red dashed) match the exact quantum results (black solid), with GDTWA slightly outperforming the Spin-PLDM result. For $\braket{x_{19}^2}$, both methods reproduce qualitative features of the exact evolution better than the other considered semiclassical techniques. 
	}
\label{FC2}
\end{figure}

Due to symmetry, all the off-diagonal elements of the electron density matrix of the two models vanish. In Fig.~\ref{fig:ThreeTwo}, we show the population of the second diabatic state of Model I. 
The GDTWA result compares fairly well to the exact quantum dynamics. It seems to underestimate the amplitude of oscillations around the mean, and reaches a long-time average that lies slightly below the exact value. The functional form seems to be better reproduced than with the Ehrenfest method, and the curve lies closer to the exact result than the curve computed using the Spin-PLDM method. Finally, the PLDM methods produces the best estimate of the long-time average, but considerably overestimates the damping of the oscillations. 
GDTWA fits the quantum result rather well at short times and has a fair performance at longer
times, though it does not outperform the other approaches in this regime. 
Figure~\ref{TT} shows the dynamics of the two tuning coordinates, $\braket{x_1}$ and $\braket{x_{6a}}$. Though GDTWA does not entirely capture the correct amplitude, it does match very well the frequency of the occurring oscillation. This behavior is similar to the Spin-PLDM method, while PLDM significantly underestimates the oscillation amplitude and the Ehrenfest method loses half a period within
about five to ten oscillations. Figure~\ref{TC2} presents the propagation of $\braket{x_1^2}$, $\braket{x_{6a}^2}$, and $\braket{x_{10a}^2}$. In general, we should not expect the linearized semi-classical methods to work reliably for such higher-order correlations. As the numerical results suggest, Spin-PLDM and GDTWA nevertheless still give qualitatively satisfactory results, while PLDM and the Ehrenfest method rather quickly accumulate uncontrolled errors. 

The relaxation dynamics of the more complex Model II is considerably more challenging for the linearized semi-classical methods because several states are involved simultaneously in the relaxation dynamics. GDTWA is the only one among the selected semi-classical methods to qualitatively correctly capture the relaxation dynamics, as seen in the diabatic populations in Figure~\ref{FiveThree}. In Figures~\ref{FT}, \ref{FT2}, and \ref{FC2}, we show the populations of the tuning coordinates as well as their diagonal second-order correlations, and the second-order diagonal correlations of the coupling coordinates, respectively. 
PLDM and the Ehrenfest method display significant deviations from the exact dynamics. In contrast, GDTWA yields surprisingly accurate predictions, for some observables even slightly but noticeably better than Spin-PLDM.

\section{Conclusions}
\label{sec:conclusions}

In this paper, we have introduced a recently developed method from the TWA family, GDTWA, to chemical non-adiabatic systems. The novelty and strength of GDTWA is to sample the electron DoF in a discrete phase space. We have also re-written the GDTWA in a form similar to the Ehrenfest method, with the aim of showcasing similarities and differences to more conventional methods. Formally, the EOMs of GDTWA are identical to fully linearized mapping approaches. By the spectral decomposition of the electron EOM, we demonstrate that the fundamental difference between GDTWA and traditional approaches is that GDTWA has two coupled correlated electron states in each single classical trajectory, and hence can be regarded as a partially linearized approach. GDTWA also accounts for an effective ZPE without an explicit ZPE parameter. Numerical benchmarks show the validity of GDTWA for non-adiabatic systems. For the two benchmark LVC models in this paper,
GDTWA displays qualitative and quantitative accuracy compared to the quantum description.
For one of the considered models, it even outperforms the Spin-PLDM, which is the only other of the considered methods to display an at least qualitative agreement for most of the considered situations. 

Various extensions of the GDTWA are in progress, namely, the coupling of the system to time-dependent electromagnetic fields and the extension of GDTWA to simulations in the adiabatic representation, which will enable, e.g., on-the-fly simulations in conjunction with usual electronic structure packages for the electronic structure. The performance of the method in such scenarios will be reported in future works. 

\section*{Acknowledgments}
We acknowledge support by Provincia Autonoma di Trento, the ERC Starting Grant StrEnQTh (Project-ID 804305), Q@TN --- Quantum Science and Technology in Trento.

\section*{Data AVAILABILITY}
The data that support the findings of this study are available within the article.

\appendix

\section{Explicit form of $\hat{\Lambda}_\mu$ with $N=2$ and $N=3$}
\label{sec:ExplicitLambdamu}
The $\hat{\Lambda}_\mu$ used in the main text form the basis of $SU(N)$, and can thus be represented as $N-1$ matrices of size $N\times N$, plus the identity matrix. 

When $N=2$, the basis elements are simply proportional to the Pauli matrices,
\begin{equation}
\begin{aligned}
    \hat{\Lambda}_1 = \frac{1}{\sqrt{2}}\left( \begin{array}{cc}
0 & 1\\
1 & 0\\
\end{array} 
\right ),\quad    \hat{\Lambda}_2 = \frac{1}{\sqrt{2}}\left( \begin{array}{cc}
0 & -i\\
i & 0\\
\end{array} 
\right ),\\ 
\hat{\Lambda}_3 = \frac{1}{\sqrt{2}}\left( \begin{array}{cc}
1 & 0\\
0 & -1\\
\end{array} 
\right ),\quad    \hat{\Lambda}_4 = \frac{1}{\sqrt{2}}\left( \begin{array}{cc}
1 & 0\\
0 & 1\\
\end{array} 
\right ).
\end{aligned}
\end{equation}

When $N=3$, they are proportional to the Gell--Mann matrices, 
\begin{equation}
\begin{aligned}
    \hat{\Lambda}_1 &= \frac{1}{\sqrt{2}}\left( \begin{array}{ccc}
0 & 1 & 0\\
1 & 0 & 0\\
0 & 0 & 0\\
\end{array} 
\right ), \,      \hat{\Lambda}_2 = \frac{1}{\sqrt{2}}\left( \begin{array}{ccc}
0 & 0 & 1\\
0 & 0 & 0\\
1 & 0 & 0\\
\end{array} 
\right ),\\ 
\hat{\Lambda}_3 &= \frac{1}{\sqrt{2}}\left( \begin{array}{ccc}
0 & 0 & 0\\
0 & 0 & 1\\
0 & 1 & 0\\
\end{array} 
\right ), \,      \hat{\Lambda}_4 = \frac{1}{\sqrt{2}}\left( \begin{array}{ccc}
0 & 0 & 1\\
0 & 0 & 0\\
1 & 0 & 0\\
\end{array} 
\right ),\\ 
\hat{\Lambda}_5 &= \frac{1}{\sqrt{2}}\left( \begin{array}{ccc}
0 & -i & 0\\
i & 0 & 0\\
0 & 0 & 0\\
\end{array} 
\right ), \,      \hat{\Lambda}_6 = \frac{1}{\sqrt{2}}\left( \begin{array}{ccc}
0 & 0 & 0\\
0 & 0 & -i\\
0 & i & 0\\
\end{array} 
\right ),\\ 
    \hat{\Lambda}_7 &= \frac{1}{\sqrt{2}}\left( \begin{array}{ccc}
1 & 0 & 0\\
0 & -1 & 0\\
0 & 0 & 0\\
\end{array} 
\right ), \,      \hat{\Lambda}_8 = \frac{1}{\sqrt{6}}\left( \begin{array}{ccc}
1 & 0 & 0\\
0 & 1 & 0\\
0 & 0 & -2\\
\end{array} 
\right ),\\ 
    \hat{\Lambda}_9 &= \frac{1}{\sqrt{3}}\left( \begin{array}{ccc}
1 & 0 & 0\\
0 & 1 & 0\\
0 & 0 & 1\\
\end{array} 
\right ).
\end{aligned}
\end{equation}

\section{Sampling of the intra-electronic correlation}

The faithful sampling for the intra-electronic correlation is crucial for the accuracy of GDTWA for the non-adiabatic dynamics. The reason is that, once there is a non-zero nuclei-electron coupling, the intra-electron correlation terms appear in the higher-order time derivatives of the EOMs. We report the detailed analysis for the diabatic basis in this appendix to show how the intra-electronic correlations affect the accuracy of GDTWA. After a lengthy but straightforward calculation, we obtain the second- and the third-order time derivative of $\lambda_\alpha$ and $\hat{\Lambda}_\alpha$,

\begin{widetext}

\begin{equation}\label{eq:stat2}
    \frac{{\rm d}^2\lambda_\mu(t)}{{\rm d}t^2} = f_{\mu\nu\xi}[\partial_{x_t}v_\nu(x_t) \frac{p_t}{m}\lambda_\xi+v_\nu(x_t)\frac{p_t}{m}f_{\xi\delta\epsilon}v_\delta(x_t)\lambda_\epsilon]\,,
\end{equation}
\begin{equation}\label{eq:quant2}
    \frac{{\rm d}^2\hat{\Lambda}_\mu(t)}{{\rm d}t^2} = f_{\mu\nu\xi}[ \frac{\partial_xv_\nu(\hat{x}_t)\hat{p}_t}{2m}\hat{\Lambda}_\xi+\frac{v_\delta(\hat{x}_t)v_\nu(\hat{x}_t)\hat{p}_t}{2m}f_{\xi\delta\epsilon}\hat{\Lambda}_\epsilon] + \rm h.c.\,,
\end{equation}
\begin{equation}\label{eq:stat3}
    \frac{{\rm d}^3\lambda_\mu(t)}{{\rm d}t^3} = f_{\mu\nu\xi}[\partial_{x_t}^2v_\nu(x_t) \frac{p_t^2}{m^2}\lambda_\xi-\partial_{x_t}v_\nu(x_t)\frac{1}{m}\partial_{x_t}v_\zeta(x_t)\lambda_\zeta\lambda_\xi+\partial_{x_t}v_\nu(x_t)\frac{p_t}{m}f_{\xi\delta\epsilon}v_\delta(x_t)\lambda_\epsilon]\,,
\end{equation}
\begin{equation}\label{eq:quant3}
    \frac{{\rm d}^3\hat{\Lambda}_\mu(t)}{{\rm d}t^3} = f_{\mu\nu\xi}[ \frac{\partial_{\hat{x}_t}^4v_\nu(\hat{x}_t)+4\partial_{\hat{x}_t}^2v_\nu(\hat{x}_t)\hat{p}_t^2}{8m^2}\hat{\Lambda}_\xi-\partial_{\hat{x}_t}v_\nu(\hat{x}_t)\frac{1}{2m}\partial_{\hat{x}_t}v_\zeta(\hat{x}_t)\hat{\Lambda}_\zeta\hat{\Lambda}_\xi+\frac{v_\delta(\hat{x}_t)\partial_{\hat{x}_t}v_\nu(\hat{x}_t)\hat{p}_t}{2m}f_{\xi\delta\epsilon}\hat{\Lambda}_\epsilon] + \rm h.c.\,,
\end{equation}
\end{widetext}
where ${\rm h.c.}$ is the Hermitian conjugate. We focus on the short time $t\sim 0$ accuracy. As for the separable initial state $\rho(0)$ the statistical average of Eq.~(\ref{eq:stat2}) is identical to the quantum expectation value of Eq.~(\ref{eq:quant2}), the GDTWA is at least accurate up to $\mathcal{O}(t^2)$. Meanwhile,
the statistical average of Eq.~(\ref{eq:stat3}) equals the quantum expectation value of Eq.~(\ref{eq:quant3}) if Eq.~(\ref{eq:Intra}), the condition of faithful statistical sampling of the initial intra-electron correlations, is fulfilled. Thus, in this case the accuracy of GDTWA is improved for the short time dynamics, as it is ensured to be exact at least up to and including $\mathcal{O}(t^3)$. 

We stress that ``intra-electron correlation'' here denotes only a feature of statistical sampling, to be distinguished from the correlation between nuclear and electronic DoFs, or the static correlation and dynamical correlation in the electronic structure theory. We illustrate how the discrete sampling fails to represent the intra-electronic correlation at the example of an explicit state without the nuclei-electron correlation. Consider the state $\ket{\Psi}=(\ket{1}+e^{i\chi}\ket{2})/\sqrt{2}$ for a two-level system, where the discrete sampling gives the probability distribution 
\begin{equation}
\begin{aligned}
p(\lambda_1=\pm\frac{1}{\sqrt{2}})=\frac{1\pm\cos{\chi}}{2},\\
p(\lambda_2=\pm\frac{1}{\sqrt{2}})=\frac{1\pm\sin{\chi}}{2},\\
p(\lambda_3=\pm\frac{1}{\sqrt{2}})=\frac{1}{2}.
\end{aligned}
\end{equation}
With an explicit calculation, we obtain  $\frac{\hat{\Lambda}_1\hat{\Lambda}_2+\hat{\Lambda}_2\hat{\Lambda}_1}{2}=0$, while 
\begin{equation}
\begin{aligned}
    \sum_{a_1,a_2}
    p(\lambda_1=a_1)p(\lambda_2=a_2)a_1 a_2=\frac{\sin{2\chi}}{4},
\end{aligned}
\end{equation}
which means the discrete sampling of this state is faithful for the intra-electron correlation only if $\chi=0$, $\pi$, or $\pm\pi/2$.

\section{Different ZPE between SM and GDTWA for block diagonal Hamiltonians}

Though SM and GDTWA have an identical dimension dependency of the ZPE, they may behave differently when the Hamiltonian is block diagonal.
Consider a simple $N\times N$ Hamiltonian with the elements $H_{kl} = 0$ for $M<k\le N$, $1\le l\le M$ and $1\le k\le M$, $M<l\le N$. The first $M$ diabatic states are decoupled from the other $N-M$ states.
Again, we only consider the initial state $\ket{1}\bra{1}$. As before, we denote the electron phase space variable of the $N\times N$ full electron system as $A_\alpha(t)$ and $B_\alpha(t)$
while the submatrix $A_\alpha(t)[1,2,\cdots,M;1,2,\cdots,M]$ is indicated as $A_\alpha^M(t)$ (and analogously for $B_\alpha$). 

Since the first $M$ diabatic states are decoupled from the others, it is also possible to sample the $M\times M$ subsystem directly. We use $\tilde{A}_\alpha^M(t)$ and $\tilde{B}_\alpha^M(t)$ to represent the electron phase space variables obtained by sampling from the $M\times M$ subsystem. It is easy to check that the initial distributions of $A_\alpha^M(0)$ and $\tilde{A}_\alpha^M(0)$ are identical. Moreover, the classical trajectories satisfy $A_\alpha^M(t)=\tilde{A}_\alpha^M(t)$ if their initial conditions are the same. Thanks to the implicit ZPE parameter of GDTWA, all the physical quantities are invariant independent of whether we use the $N\times N$ full electron system or the $M\times M$ subsystem.

The above arguments become much more subtle for the SM approach with the dimension dependent ZPE parameter. The initial distribution of $B_\alpha^M(0)$ and $\tilde{B}_\alpha^M(0)$ become different, as do the classical trajectories, even when the same initial conditions are applied. This difference may affect the observables, though it is difficult to give a general statement under which circumstances this is the case.

\bibliography{main}

\end{document}